%% file: MesonPLB.tex
\documentclass[preprint,5p,twocolumn,10pt]{elsarticle}
\usepackage[pdftex,hypertexnames=false,linktocpage=true]{hyperref}
\hypersetup{colorlinks=true,linkcolor=red,anchorcolor=darkgreen,citecolor=green!50!blue,filecolor=darkgreen,urlcolor=green,bookmarksnumbered=true,pdfview=FitB}

\usepackage{lipsum}
\usepackage{mathtools,cuted}

\usepackage{amsmath}
\usepackage{graphicx}

\usepackage{array}
\usepackage{booktabs}
\usepackage{arydshln}
\usepackage{multirow}

\newcommand{\ie}{\textit{i.e.}}

\input{preamble_math_physics.tex}

\usepackage{braket}

\graphicspath{{../../plots/}}

\begin{document}

\begin{frontmatter}
%\title{Strange mesons with one dynamical gluon in a light front approach}
\title{Strange mesons with one dynamical gluon: A light-front  approach}

\author[imp,djl,ucas,ashz]{Jiangshan Lan\corref{cor1}}
\ead{jiangshanlan@impcas.ac.cn}

\author[imp,ucas]{Jialin Chen}
\ead{chenjialin@impcas.ac.cn}

\author[imp,ucas]{Zhimin Zhu\corref{cor1}}
\ead{zhuzhimin@impcas.ac.cn}

\author[imp,ucas]{Chandan Mondal\corref{cor1}}
\ead{mondal@impcas.ac.cn}

\author[imp,djl,ucas]{Xingbo Zhao}
\ead{xbzhao@impcas.ac.cn}

\author[iowa]{James P. Vary}
\ead{jvary@iastate.edu}

% \collab[BLFQ]{BLFQ Collaboration}
\author[]{\\\vspace{0.2cm}(BLFQ Collaboration)}
% \collaboration{BLFQ Collaboration}
\date{\today}

\address[imp]{Institute of Modern Physics, Chinese Academy of Sciences, Lanzhou, Gansu 730000, China}
\address[djl]{Advanced Energy Science and Technology Guangdong Laboratory, Huizhou, Guangdong 516000, China}
\address[ucas]{School of Nuclear Physics, University of Chinese Academy of Sciences, Beijing  100049, China}

\address[ashz]{ Affiliated School of Huizhou University, Huizhou, Guangdong  516001, China}
\address[iowa]{Department of Physics and Astronomy, Iowa State University, Ames, Iowa 50011, USA}

\cortext[cor1]{Corresponding author}

\begin{abstract}

We obtain the mass spectra of strange mesons using a light-front quantized Hamiltonian with Quantum Chromodynamics (QCD) input, incorporating quark-antiquark and quark-antiquark-gluon Fock components, along with a three-dimensional confinement.  We work within the basis light-front quantization framework. The resulting eigenvectors can simultaneously describe the kaon's electromagnetic form factor, decay constant, distribution amplitude, and quark and gluon distribution functions under QCD scale evolution. Using the obtained kaon parton distribution functions (PDFs), supplemented by established nuclear PDFs, we also predict the kaon-nucleus-induced Drell-Yan cross section, which is expected to be measured soon by COMPASS++/AMBER at CERN.

\end{abstract}
\end{frontmatter}

\section{Introduction}
Light strange mesons, such as kaons~\cite{Rochester:1947mi}, are a class of hadrons composed of a strange quark and a lighter up or down quark according to the quark model~\cite{Gell-Mann:1964ewy}. These mesons play a critical role in understanding the strong force, as they serve as ideal systems for studying the interplay between light and strange quarks within Quantum Chromodynamics (QCD). Kaons, in particular, are key to exploring symmetry breaking in the Standard Model, especially in relation to charge-parity (CP) violation~\cite{Woods:1988za,NA31:1993tha,E731:1993niw}, which has profound implications for understanding the matter-antimatter asymmetry in the universe. The mass spectra, decay constants, and electromagnetic properties of strange mesons provide essential data for testing theoretical models, including lattice QCD and light-front quantization techniques. Moreover, strange mesons are pivotal in high-energy physics experiments, such as those conducted at CERN, where they help probe the structure of nucleons and nuclei, offering insights into the behavior of partons at different scales.

Electromagnetic form factors (EMFFs) and parton distribution functions (PDFs) provide key insights into the internal structure of hadrons. The Fourier transform of FFs reveals spatial distributions, such as charge and magnetization. Deep inelastic scattering (DIS) and Drell-Yan processes further probe hadron structure by resolving individual quarks, allowing the extraction of PDFs that describe the distribution of longitudinal momentum and polarization among the constituents. To date,  experimental investigations of the charged kaon EMFFs cover a spacelike region that extends to $Q^2\sim 2.35$ GeV$^2$~\cite{Amendolia:1986ui,Dally:1980dj,Carmignotto:2018uqj}, where $Q^2$ is the squared momentum transfer in the process. Upcoming experiments are expected to extend these measurements beyond $Q^2\sim 30$ GeV$^2$~\cite{Arrington:2021biu,Chen:2020ijn,Proceedings:2020fyd}. On the other hand, experimental data on the light-quark PDF of kaons is available in the form of the ratio of the up quark valence PDF in kaons to that in pions~\cite{Conway:1989fs,NA3:1983ejh} from the NA-003 experiment at CERN. However, the strange quark and gluon distributions in kaons have yet to be determined.

 Several future experiments have the potential to obtain new  data on the internal structure of the kaons. The COMPASS++/AMBER at CERN plans to investigate the valence-quark distributions of the kaon by $K^{-}$-induced Drell-Yan measurements and to investigate the gluon distributions of the kaon by $K^{-}$-induced $J/\Psi$ production on a $^{12}C$ target \cite{Adams:2018pwt}. The next generation experimental facilities, Electron-ion collider (EIC) and Electron-ion collider in China (EicC) experiments will also focus on the measurements of the quark and gluon distribution functions of the kaon~\cite{Accardi:2012qut,AbdulKhalek:2021gbh,Arrington:2021biu,Anderle:2021wcy}.

Theoretical studies have made significant strides in understanding kaon structure using various QCD-inspired models and approaches. These include the constituent quark model~\cite{Londergan:1996vh,Gluck:1997ww}, chiral constituent quark model ($\chi$QM)~\cite{Watanabe:2018qju}, gauge-invariant nonlocal chiral quark model (NL$\chi$QM)~\cite{Hutauruk:2023ccw}, Nambu–Jona-Lasinio (NJL) model~\cite{Hutauruk:2021kej}, Dyson-Schwinger/Bethe-Salpeter equations (DSE/BSE)~\cite{Roberts:2021nhw,Xu:2018eii,Nguyen:2011jy,Bednar:2018mtf}, QCD instanton liquid model (ILM)~\cite{Liu:2023yuj}, light-cone quark model (LCQM)~\cite{Kaur:2020vkq}, basis light front quantization with color-singlet Nambu–Jona-Lasinio interactions (BLFQ-NJL) model~\cite{Lan:2019vui}, and lattice QCD~\cite{PhysRevD.103.014516,Alexandrou:2021mmi}. However, these models/approaches diverge on the behavior of the up quark distribution at large longitudinal momentum fractions ($x$) and the ratio of momentum carried by the strange antiquark versus the up quark~\cite{Watanabe:2018qju,Hutauruk:2023ccw}. The predictions for the gluon distribution also vary across models~\cite{Londergan:1996vh,Hutauruk:2023ccw,Salas-Chavira:2021wui}.

BLFQ offers a nonperturbative framework for solving relativistic many-body bound state problems in quantum field theory~\cite{Vary_2010}. This method determines light-front wave functions (LFWFs) and mass-squared eigenvalues by diagonalization of a Hamiltonian expressed in a Fock sector expansion and has been successfully applied to both QED~\cite{Honkanen:2010rc,Zhao:2014xaa,Wiecki:2014ola,Hu:2020arv,Nair:2022evk} and QCD bound systems~\cite{Li:2017mlw,Jia:2018ary,Lan:2019vui,Lan:2019rba,Lan:2019img}. Notably, BLFQ has investigated the light meson system with a higher Fock sector containing a dynamical gluon, providing a good description of the pion~\cite{Lan:2021wok,Zhu:2023lst}. 

In this work, we investigate strange mesons using the BLFQ framework, extending the Fock sectors to include both $|q\bar{q}\rangle$ and $|q\bar{q}g\rangle$ components, incorporating vector coupling vertex interactions, and a three-dimensional confinement in the $|q\bar{q}\rangle$ sector. For the strange meson system, we introduce two additional parameters: the model strange quark mass and the vertex interaction mass, while keeping other parameters consistent with those used for light mesons~\cite{Lan:2021wok}. We fit the experimental masses of the strangeonium states $\phi(1020)$ and $f_1(1420)$~\cite{ParticleDataGroup:2022pth} to determine these parameters. Our approach yields a suitable description of strange meson and strangeonium mass spectra. We also evaluate the kaon EMFF, distribution amplitudes (DAs), and decay constant, calculate and evolve the PDFs of the kaon using QCD scale evolution, and predict the kaon-nucleus induced Drell-Yan  cross section.

%====================================================
\section{Mass spectra and LFWFs of strange mesons}
%====================================================
The meson state in light-front (LF) field theory can be obtained by solving the LF stationary Schr\"{o}dinger equation~\cite{Kogut:1969xa,Brodsky:1997de},
\begin{equation}
    P^-P^+|{\Psi}\rangle=M^2|{\Psi}\rangle,\label{eigenequation}
\end{equation}
where $P^\pm=P^0 \pm P^3$. $P^-$ and $P^+$ represent the LF Hamiltonian and the longitudinal momentum, respectively. The eigenvalue $M^2$ is the mass squared of the meson state. At fixed LF time, $x^+=x^0+x^3$, a meson state can be expanded into various quark ($q$), antiquark ($\bar q$), and gluon $(g)$ Fock components in the Fock space. With the Fock sector expansion, the meson state can be schematically expressed as 
\begin{align}
    |\Psi\rangle=\psi_{(q\bar{q})}|q\bar{q}\rangle+\psi_{(q\bar{q}g)}|q\bar{q}g\rangle+\dots\, , \label{Fockexpansion}
\end{align}
where the LFWFs $\psi_{(\dots)}$ correspond to the projection of the meson state to the Fock sectors $|\cdots\rangle$~\cite{Brodsky:1997de}. In this study, we apply the Fock sector truncation at the level of three constituent particles, \ie, the mesons are described within two different Fock sectors $|q\bar{q}\rangle$ and $|q\bar{q}g\rangle$. 

The current LF Hamiltonian we use contains the LF QCD Hamiltonian, $P^-_{\rm QCD}$, and the confinement, $P^-_{\rm{C}}$, \ie, $P^-= P^-_{\rm QCD} +P^-_{\rm{C}}$~\cite{Lan:2021wok}. With one dynamical gluon, the LF QCD Hamiltonian in the LF gauge $A^+=0$ reads~\cite{Brodsky:1997de,Lan:2021wok}
\begin{align}
    P_{\rm QCD}^-=& \int \mathrm{d}x^-\mathrm{d}^2 x^{\perp} \Bigg[\frac{1}{2}\bar{\psi}\gamma^+\frac{m_{0}^2+(i\partial^\perp)^2}{i\partial^+}\psi\nonumber\\
    & -\frac{1}{2}A_a^i(i\partial^\perp)^2 A^i_a +g_s\bar{\psi}\gamma_{\mu}T^aA_a^{\mu}\psi \nonumber\\
    &+\frac{1}{2}g_s^2\bar{\psi}\gamma^+T^a\psi\frac{1}{(i\partial^+)^2}\bar{\psi}\gamma^+T^a\psi \Bigg],\label{eqn:PQCD}
\end{align}
where $\psi$ and $A^\mu$ are the quark and gluon fields, respectively. $T^a$ is the half Gell-Mann matrix, $T^a=\lambda^a/2$, and $\gamma^+=\gamma^0+\gamma^3$, where $\gamma^\mu$ represents the Dirac matrix. The first two terms in Eq.~(\ref{eqn:PQCD}) are the kinetic energies of the quark and the gluon, while the last two terms describe their interactions with coupling constant $g_s$. $m_{0}$ and $m_g$ are the bare mass of quarks and the model gluon mass, respectively. Using the Fock sector dependent renormalization scheme~\cite{Karmanov:2008br,Li:2015iaw}, we introduce a mass counter term, $\delta m_{q}= m_0 -m_{q}$, in the leading Fock sector to regularize the quark self-energy. Here, $m_{q}$ is the renormalized quark mass. Regarding the vertex interaction, we introduce a different quark mass $m_f$ referring to Ref.~\cite{Glazek:1992aq}. 

The confinement in the leading Fock sector includes transverse and longitudinal confining potentials~\cite{Li:2015zda,Lan:2021wok},
\begin{equation}
  \begin{split}\label{eqn:Hc}
  &P_{\rm C}^-P^+=\kappa^4\left\{x(1-x) \vec{r}_{\perp}^2-\frac{\partial_{x}[x(1-x)\partial_{x}]}{(m_q+m_{\bar{q}})^2}\right\},
  \end{split}
\end{equation}
where the parameter $\kappa$ is the strength of the confinement, and $\vec r_{\perp}=\sqrt{x(1-x)}(\vec r_{\perp q}-\vec r_{ \perp\bar{q}})$ represents the holographic variable \cite{Brodsky:2014yha}. Whereas the confinement in the $|q\bar{q}g\rangle$ sector is achieved by the presence of the massive gluon and the cutoff of the BLFQ basis functions discussed below. 

In this work, we follow BLFQ~\cite{Vary_2010,Vary:2009gt,Zhao:2013cma,Wiecki:2014ola,Zhao:2014xaa,Li:2015zda,Lan:2021wok,Xu:2021wwj,Kaur:2024iwn}, a non-perturbative computational framework, to solve the LF eigenvalue problem, Eq.~(\ref{eigenequation}). Each Fock sector, $|\cdots\rangle$ in Eq.~(\ref{Fockexpansion}), is the direct product of the single-particle states $|\alpha\rangle=\otimes_i|\alpha_i\rangle$. In the BLFQ framework, the Discretized Light Cone Quantization basis~\cite{Brodsky:1997de} is used to describe the longitudinal degrees of freedom of the single-particle states. The longitudinal motion of a fermion (boson) is confined to a one-dimensional box of length $2L$ with antiperiodic (periodic) boundary conditions. So the longitudinal momentum is then defined as $p^+=\frac{2\pi}{L}k$, where the longitudinal quantum number $k$ for boson (fermion) is an integer (half-integer). Here, we neglect the zero modes for boson. Meanwhile, the transverse degrees of freedom are described by the two-dimensional harmonic-oscillator (2D-HO) basis functions, $\Phi_{nm}(\vec{p}_{\perp},b)$, and $b$ is the HO scale. $n$ and $m$ denote the radial and the angular quantum numbers, respectively. Each single-particle basis state contains four quantum numbers $|\alpha_i\rangle=|k_i,n_i,m_i,\lambda_i\rangle$, where the quantum number $\lambda$ denotes the light-front helicity. In Fock sectors that permit multiple color-singlet states (i.e., beyond the Fock sectors considered here), an additional label is required to distinguish each color-singlet state. In addition, our many-body basis states have well-defined total angular momentum projections $M_J=\sum_i (m_i+\lambda_i)$. 

For numerical calculations, we truncate the infinite basis space by respectively introducing two truncation parameters $K$ and $N_{\mathrm{max}}$ in longitudinal and transverse directions, such that $K=\sum_i k_i$ and $\sum_i(2n_i+|m_i|+1)\leq N_{\mathrm{max}}$. The truncation $K$ denotes the resolution of PDFs, \ie, the longitudinal momentum fraction $x_i=p^+/P^+=k_i/K$. The truncation $N_{\mathrm{max}}$ enables factorization of the transverse center of mass motion \cite{Wiecki:2014ola,Zhao:2014xaa} and implicitly acts as the infrared (IR) cutoff, $\lambda_{\rm IR} \simeq b/\sqrt{N_{\rm max}}$, and the ultraviolet (UV) cutoff, $\Lambda_{\rm UV} \simeq b\sqrt{N_{\rm max}}$, in the  momentum space~\cite{Zhao:2014xaa}. 

By solving the eigenvalue equation, Eq.~(\ref{eigenequation}), we obtain the mass spectra $M^2$ and the corresponding LFWFs in momentum space,
\begin{align}
  &\Psi^{M_J}_{\mathcal{N},\{\lambda_i\}}({\{x_i,\vec{p}_{\perp i}\}})\nonumber\\
  &=\sum_{ \{n_i m_i\} }\psi^{M_J}_{\mathcal{N}}({\{\alpha}_i\})\prod_{i=1}^{\mathcal{N}}  \Phi_{n_i m_i}(\vec{p}_{\perp i},b)\,,
\label{eqn:wf}
\end{align}
where $\psi^{M_J}_{\mathcal{N}=2}(\{\alpha_i\})$ and $\psi^{M_J}_{\mathcal{N}=3}(\{\alpha_i\})$ are the components of the eigenvectors associated with the Fock sectors $|q\bar{q}\rangle$ and $|q\bar{q}g\rangle$, respectively.

\begin{table}[htp]
  \caption{The parameters, $\{m_q,m_{f_q},m_g,b,\kappa,g_s\}$, are identical to those in Ref.~\cite{Lan:2021wok}. The strange quark masses, $m_s$ and  $m_{f_s}$, are determined by fitting $\phi(1020)$ and $f_1(1420)$ from experiments. All quantities are in units of GeV except $g_s$.}
  \vspace{0.15cm}
  \label{para}
  \centering
  \begin{tabular}{cccccccc}
    \hline\hline
         $m_q$ &${m}_{f_q}$ & $m_g$ &$b$ &$\kappa$ &$g_s$ &$m_s$ &$m_{f_s}$ \\        
    \hline 
        0.39 &5.69 & 0.60 &0.29 &0.65 &1.92 & 0.55 &7.13 \\       
    \hline\hline
  \end{tabular}
\end{table}

In our previous work~\cite{Lan:2021wok}, we fitted the unflavored meson mass spectra to fix the model parameters, $\{m_q,m_{f_q},m_g,b,\kappa,g_s\}$, summarized in Table \ref{para}, with the truncations $\{N_{\text{max}},K\}=\{14,15\}$. For strange mesons, we adopted the same parameters as in Ref.~\cite{Lan:2021wok} except for the two additional parameters: the model strange quark masses in kinetic energies $m_s$ and vertex interactions $m_{f_s}$, shown in Table \ref{para}. 
Here, we select two established strangeonium states, $\phi(1020)$ and $f_1(1420)$, in the Particle Data Group 2022 (PDG 22)~\cite{ParticleDataGroup:2022pth} for fitting the two additional parameters.

%===================================
\begin{figure}
 \centering
    % \begin{center}
    \includegraphics[width=0.97\linewidth]{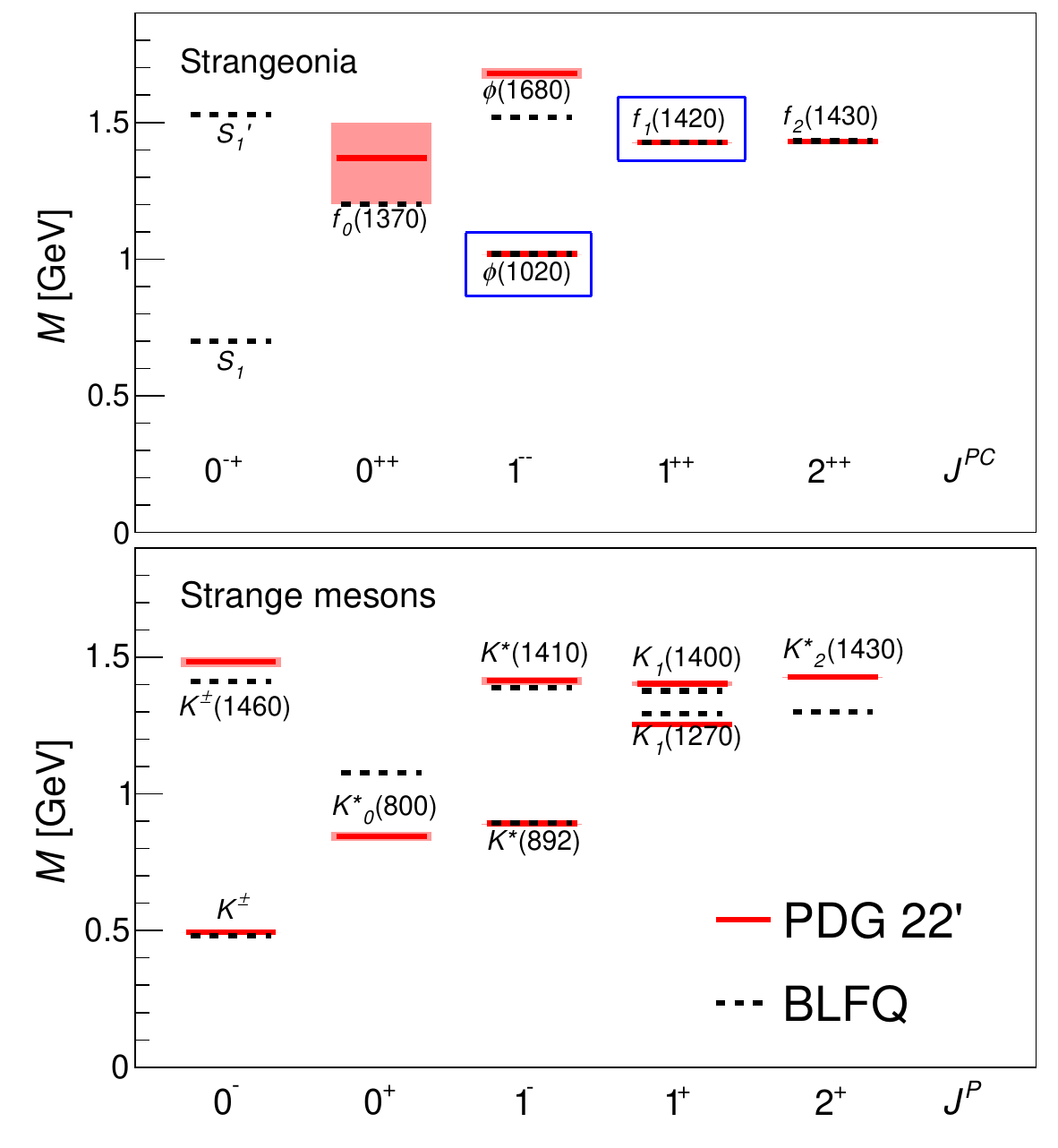}
    \caption{The mass spectra of strangeonia (upper) and strange mesons (lower). 
    % The two states with the blue box are the two established strangeonium states, $\phi(1020)$ and $f_1(1420)$, we select for fitting.
    The two states with the blue box represent the two established strangeonium states, $\phi(1020)$ and $f_1(1420)$, which we select for fitting.
    Our results (Black-dashed bars) are compared with the experimental data taken from PDG 22 (red-solid bars)~\cite{ParticleDataGroup:2022pth}.}
    \label{fig_mass}
    % \end{center}
\end{figure}
%====================================

Figure~\ref{fig_mass} shows the mass spectra for strangeonia and strange mesons, comparing them with experimental data from PDG 22~\cite{ParticleDataGroup:2022pth}. Our results for the strangeonia spectrum align well with the experimental data. Additionally, our calculations predict two strangeonia states, $S_1$ and $S_1^\prime$, which have not been observed experimentally. For strange mesons, including 
$K^\pm$, $K^\pm(1460)$, $K^{*}(892)$, $K^{*}(1410)$, $K_1(1270)$, and $K_1(1400)$, our results agree well with experimental data. However, the states $K^{}_0(800)$ and $K^*{}_2(1430)$ 
show some deviation. Notably, by adjusting only the strange quark mass, we achieve a mass spectrum for strange mesons that closely matches experimental data, particularly for the kaon, $K^\pm$.

%======================================
\section{Kaon EMFF and decay constant}
%======================================
The EMFF  is one of the most important QCD quantities related to the transition from the non-perturbative to the perturbative regime~\cite{MartinContreras:2021yfz}.
The kaon EMFF $F(Q^2)$ is defined by the following matrix element~\cite{Brodsky:2007hb},
\begin{align}
	\langle  K^\pm (p^{\prime})| J_{\text{EM}}^\mu (0) | K^\pm (p) \rangle = (p + p^{\prime})^\mu F(Q^2),\label{EMFF_matrix_element}
\end{align}
where 
$p$ and $p^\prime$ represent the initial and final kaon momenta, respectively, with $Q^2=-(p^{\prime}-p)^2$ denoting the momentum transfer. The electromagnetic current is given by $J_{\mathrm{EM}}^\mu(z)=\sum_f e_f\bar{q}(z) \gamma^\mu q(z)$, where $f$ includes $u$ and $s$ quarks, and $e_f$ represents the quark electric charge.

Taking $\mu=+$, the matrix element in Eq.~(\ref{EMFF_matrix_element}) can be computed by expanding the initial and final meson states with the Fock sector expansion in Eq.~(\ref{Fockexpansion}). 
Using the frame, where $q^+=(p^{\prime}-p)^+=0$, the EMFF can be expressed through the meson LFWFs using the Drell-Yan-West formula:
\begin{align}
    F(Q^2)=&\sum_{\mathcal{N},f,\lambda_i} e_f \int_{\mathcal{N}} \Psi^{M_J=0*}_{\mathcal{N},\{\lambda_i\}}({\{x_i,\vec{p}^{\,\prime}_{\perp i}\}})\Psi^{M_J=0}_{\mathcal{N},\{\lambda_i\}}({\{x_i,\vec{p}_{\perp i}\}}), 
    \label{eqn:ff_q}
\end{align}
where $\int_{\mathcal{N}}\equiv \int\frac{\prod_{i=1}^\mathcal{N}\mathrm{d}x_i\mathrm{d}^2\vec{p}_{\perp i}}{(2(2\pi)^3)^{\mathcal{N}-1}}\delta(1-\sum x_j)\delta^2(\sum \vec{p}_{\perp j})$, and for a struck parton $\vec{p}^{\,\prime}_{\perp i}=\vec{p}_{\perp i}+ (1-x_i)\vec{q}_\perp$, whereas $\vec{p}^{\,\prime}_{\perp i}=\vec{p}_{\perp i}- x_i \vec{q}_\perp$ for each spectator. The subscript $\mathcal{N}=2\,,3$ corresponds to the Fock sector $|q\bar{q}\rangle$ and $|q\bar{q}g\rangle$, respectively. In the $q^+=0$ frame, the momentum transfer is purely in the transverse direction, \ie,  $Q^2=\vec{q}^{\,2}_{\perp}$. The electric charge $e_u(e_{{s}})=\frac{2}{3}(-\frac{1}{3})$ for the up (strange) quark, while $e_g=0$ for the gluon. 

%======================================
\begin{figure}[htp]
 \centering
    \includegraphics[width=0.97\linewidth]{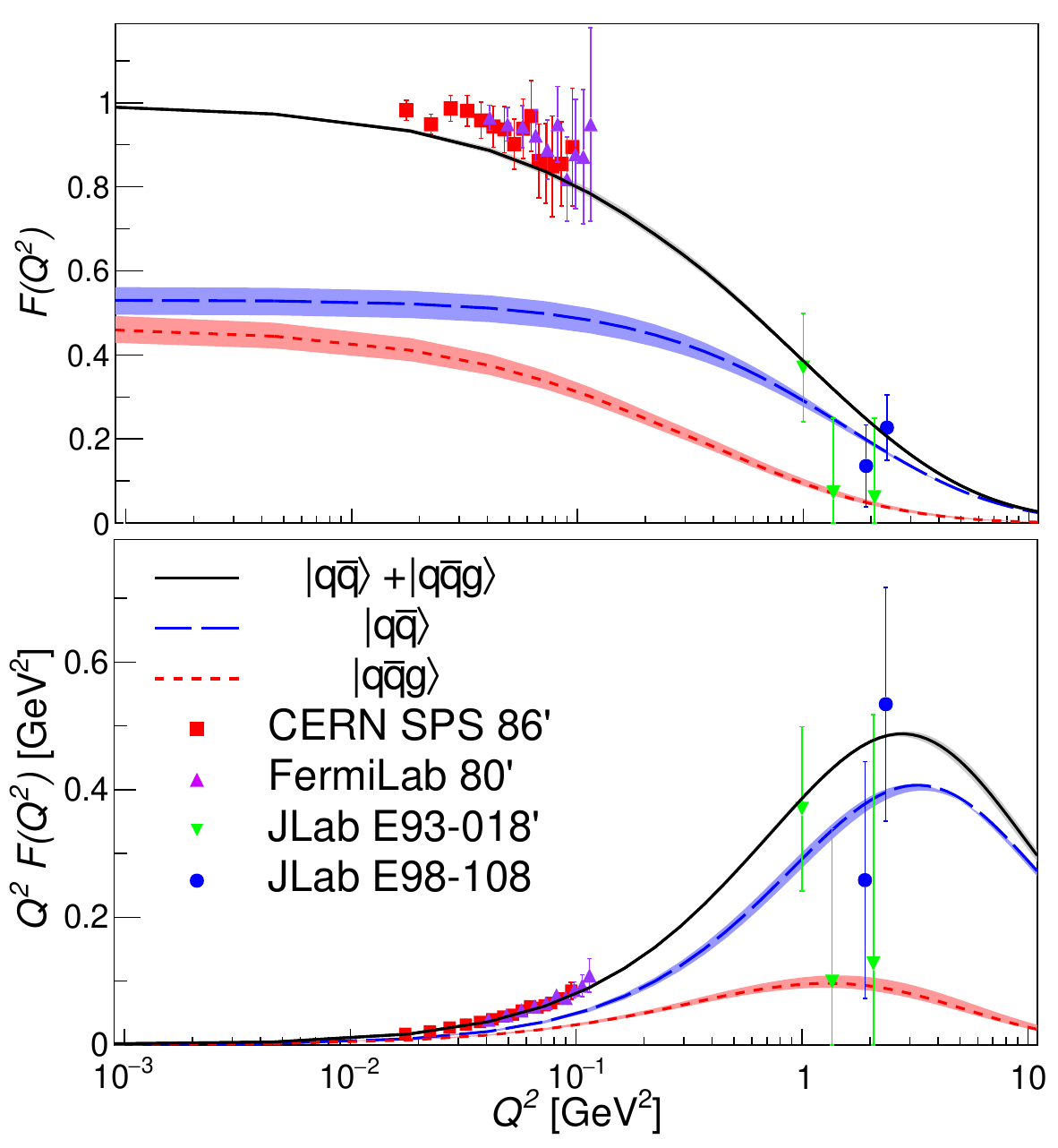}
    \caption{Our prediction for the kaon EMFF: $F(Q^2)$ (upper) and $Q^2F(Q^2)$ (lower) compared with
the experimental data~\cite{Amendolia:1986ui,Dally:1980dj,Carmignotto:2018uqj}. The blue dashed and red dotted lines represent the contributions from the $|q\bar{q}\rangle$ and $|q\bar{q}g\rangle$ Fock components, respectively, while the black line shows the total EMFF.
% The uncertainties are quantified by comparing the results at different coupling constants.
The uncertainties are indicated by light shaded bands and are based on a $\pm 10\%$ variation in $g_s$.
    }
    \label{fig:EMFF}
    % \end{center}
\end{figure}
%========================================

Figure~\ref{fig:EMFF} presents our BLFQ results for the kaon EMFF, depicted as $F(Q^2)$ and $Q^2F(Q^2)$ as functions of $Q^2$, including the bands that reflect our $g_s$ uncertainty of 10\%.
We compare our predictions with experimental data~\cite{Amendolia:1986ui,Dally:1980dj,Carmignotto:2018uqj}. The blue dashed and red dotted curves show the contributions from the $|q\bar{q}\rangle$ and $|q\bar{q}g\rangle$ Fock sectors, respectively, with the black curve representing the total EMFF. Note that the contributions from the  $ |q\bar{q}\rangle $ and $ |q\bar{q}g\rangle $ states at $Q^2=0$, $ F_{q\bar{q}}(0) $ and $ F_{q\bar{q}g}(0) $, respectively, must satisfy $ F_{q\bar{q}}(0) + F_{q\bar{q}g}(0) = F(0) = 1 $. This implies that the upper boundary of the blue band corresponds to the lower boundary of the red band, and conversely, the lower boundary of the blue band corresponds to the upper boundary of the red band.

Our results align reasonably well with experimental data in the low $Q^2$ regime. The contribution to the EMFF from the $|q\bar{q}g\rangle$ Fock sector decreases more rapidly and has negligible influence in the large $Q^2$ domain.
It is important to note that the truncation choice, $N_{\rm{max}}=14$, corresponds to a UV regulator $\Lambda_{\rm{UV}}\sim b\sqrt{N_{\rm{max}}}\approx 1\,\rm{GeV}$, indicating that our predictions are most reliable in the low $Q^2$ region. Overall, our predictions show good agreement with the available data.

The decay constant is an important parameter in particle physics, particularly in determining the strength of a meson's leptonic decays. For the kaon as a pseudoscalar meson, its decay constant, $f_K$, is directly linked to the matrix element of the axial current between the kaon state and the vacuum,
\begin{equation}
    \langle 0|\bar q(0)\gamma^\mu\gamma_5q(0)|K^\pm(p)\rangle=ip^\mu f_K.
\end{equation}
We obtain $f_K$ solely from the $|q \bar{q}\rangle$ Fock component, as the $|q \bar{q}g\rangle$ component does not contribute due to the structure of the matrix element at the leading order~\cite{ParticleDataGroup:2022pth}.  We predict $f_K^{\rm{BLFQ}}=156.9$ MeV, which is close to the experimental value of $f_K^{\rm{Exp.}}=155.6\pm 0.4$ MeV~\cite{ParticleDataGroup:2022pth}.

\section{Kaon DAs and PDFs}
%==================
LFWFs offer unique insights into light-cone distributions obtained by integrating out the transverse momentum. Among these distributions, the DAs and PDFs govern exclusive and inclusive processes at large momentum transfer, respectively~\cite{Lepage:1980fj}. In the LFWF framework, the leading-twist DA for the kaon is expressed as
\begin{align}
\frac{f_K}{2\sqrt{2N_c}}&\phi_K^i(x,\mu)=\int_{\mathcal{N}=2} \Big[\Psi^{M_J=0}_{\mathcal{N}=2,\{\uparrow\downarrow\}}({\{x_i,\vec{p}_{\perp i}\}})\nonumber\\
&-\Psi^{M_J=0}_{\mathcal{N}=2,\{\downarrow\uparrow\}}({\{x_i,\vec{p}_{\perp i}\}})\Big]\frac{1}{\sqrt{2}}\delta(x-x_i),
\end{align}
where $N_c=3$ and the DA is normalized to unity,
\begin{align}
\int_0^1 \phi_K^i(x,\mu)\mathrm{d}x=1.
\end{align}

%==============================================
\begin{figure}[htp]
\centering
    \includegraphics[width=0.97\linewidth]{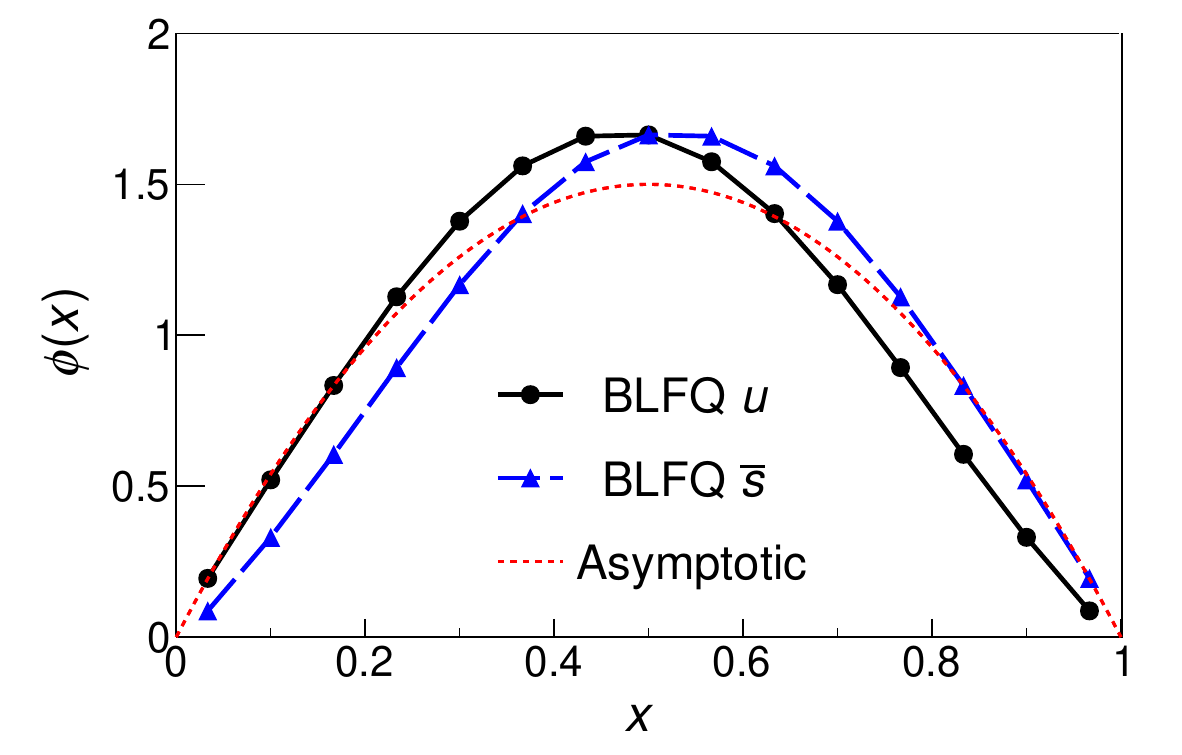}
    \caption{Our results for kaon's DAs compared with the asymptotic profile (red-dotted line). The  black-solid and blue-dashed lines with markers represent the DAs of $u$ ans $\bar{s}$ quarks in the kaon, respectively. 
    }
    \label{fig_DA}
    % \end{center}
\end{figure}
%===================

Figure~\ref{fig_DA} illustrates our results for the kaon DAs compared to the asymptotic form: $\phi_{\rm asy}(x)=6x(1-x)$~\cite{Lepage:1980fj,Lepage:1979zb,Efremov:1979qk}. Due to flavor symmetry breaking caused by the mass difference between the $\bar{s}$ and $u$ quarks, the peak of $\phi_{K}^{u/\bar{s}}(x,\mu)$ shifts away from $x=0.5$. While the width of our kaon DA closely resembles the asymptotic profile, it notably differs from the broader and flatter DA in Ref.~\cite{Roberts:2021nhw}.

The PDF represents the probability of finding a collinear parton carrying a momentum fraction $x$. Using our LFWFs, the valence quark (antiquark) and gluon PDFs in the kaon are expressed as
\begin{align}
    f_i(x)=&\sum_{\mathcal{N},\lambda_i}  \int_{\mathcal{N}}  \left|\Psi^{M_J=0}_{\mathcal{N},\{\lambda_i\}}({\{x_i,\vec{p}_{\perp i}\}})\right|^{2}\delta(x-x_i), 
    \label{eqn:pdf_q}
\end{align}
%where the integration measure is absorbed into the symbol $\int_{\mathcal{N}}$, as defined below Eq.~(\ref{eqn:ff_q}). Here, 
where $i\equiv u,\,\bar{s},\,g$ for $K^+$ and $i=s,\,\bar{u},\,g$ for $K^-$, respectively. At our model scale, the valence quark (antiquark) PDFs are normalized as $\int_0^1 f_{q/\bar{q}} (x){\rm d}x = 1$, and along with the gluon PDF, they satisfy the momentum sum rule, $\int_0^1 \sum_i x f_i(x) {\rm d}x = 1$.

We numerically solve the next-to-next-to-leading order (NNLO) Dokshitzer-Gribov-Lipatov-Altarelli-Parisi (DGLAP) equations~\cite{Dokshitzer:1977sg,Gribov:1972ri,Altarelli:1977zs} using the higher-order perturbative parton evolution toolkit~\cite{Salam:2008qg} to evolve our PDFs from the model scale ($\mu_0^2$) to a higher scale ($\mu^2$). In our previous work, we determined $\mu_0^2$ for the pion by matching the evolved valence quark and antiquark first moments to the global QCD analysis, resulting in $\mu_{0\pi}^2 = 0.34 \pm 0.03$ GeV$^2$~\cite{Lan:2021wok}. We determine our model scale, $\mu_{0K}^2 = 0.42 \pm 0.04$ GeV$^2$, for the kaon by requiring the evolved result to fit the available data for $u_v^K/u_v^{\pi}$ from the CERN-NA3 experiment~\cite{NA3:1980fhh}. The $\chi^2$ per degrees of freedom (d.o.f.) for the ratio $u_v^K/u_v^{\pi}$ is found to minimize at 1.74.

\begin{figure}[htp]
\centering
    \includegraphics[width=0.97\linewidth]{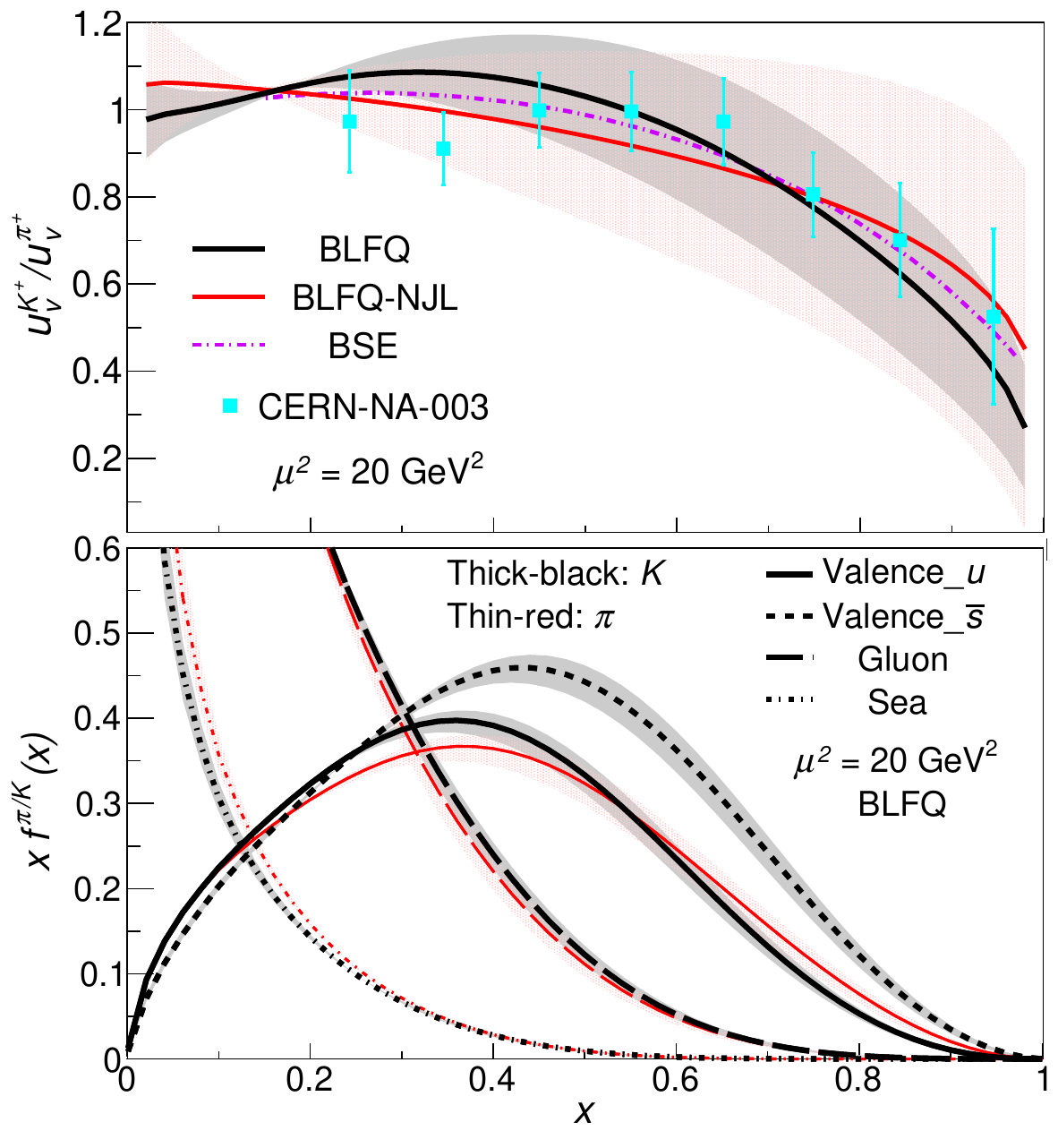}
    \caption{The PDFs of the kaon. Upper panel: the ratio of the up quark distribution in the kaon to that in the pion,  $u_v^K/u_v^{\pi}$, compared with the data from the CERN-NA3 experiment \cite{NA3:1980fhh}. Our result (black line with gray band) is compared with   the results from the BLFQ-NJL model (red line with light red band)~\cite{Lan:2019vui} and BSE approach (purple dashed line)~\cite{Nguyen:2011jy}. The gray error band corresponds to the relative errors due to the QCD evolution from the initial scales $\mu_{0K}^2=0.42 \pm 0.04 {\rm GeV}^2$ and $\mu_{0\pi}^2=0.34 \pm 0.03 {\rm GeV}^2$.
    Lower panel: comparison of the kaon PDFs (black lines) with the pion PDFs (red lines) in our BLFQ approach~\cite{Lan:2021wok} at the scale $\mu^2=20 {\rm GeV^2}$.
    }
    \label{fig_pdf}
    % \end{center}
\end{figure}

Figure~\ref{fig_pdf} presents our results for the kaon PDFs. In the upper panel, we compare our $u_v^K/u_v^{\pi}$ ratio with experimental data from the CERN-NA3 experiment~\cite{NA3:1980fhh}, along with previous results from a light-front effective Hamiltonian approach based on a valence Fock representation~\cite{Lan:2019vui} and the DSE approach~\cite{Nguyen:2011jy}. The error bands in our evolved distributions reflect a $\pm 10\%$ uncertainty in our model scales for both the kaon and pion. Our results align well with the experimental data and theoretical predictions, showing that the $u_v^K/u_v^{\pi}$ ratio decreases as $x$ increases.

This trend is further explained in the lower panel of Fig.~\ref{fig_pdf}, where we compare the valence quark distributions of the kaon and pion. For $x > 0.6$, the up quark distribution in the kaon is lower than in the pion, while for $0.2 < x < 0.6$, the up quark PDF in the kaon slightly exceeds that of the pion. These observations closely match the NA3 experimental data. Further analysis of our BLFQ results shows that, at large $x$, the kaon valence quark distributions fall off as $(1-x)^\beta$, with the up quark characterized by $\beta_u = 2.24$ and the strange quark by $\beta_s = 1.72$ at the scale $\mu^2 = 20 ~ {\rm GeV^2}$. The strange quark distribution in the kaon is significantly higher than that of the up quark at large $x$, indicating that the heavier quark carries more longitudinal momentum, as expected. Additionally, the gluon and sea quark distributions in the kaon are similar to those in the pion.

 %=======================================
%===============================================

We calculate the first moments of the PDFs, defined as $\langle x\rangle_i=\int_0^1 x f_i(x) {\rm d}x$. At the scale $\mu^2=20 ~{\rm GeV^2}$, the first moments of the kaon PDFs in our BLFQ approach are: valence up quark $\langle x\rangle_{u}=0.21\pm0.01$, valence strange antiquark $\langle x\rangle_{\bar{s}}=0.26\pm0.01$, sea quarks $\langle x\rangle_{\rm sea}=0.11\pm0.01$, and gluon $\langle x\rangle_{\rm gluon}=0.42\pm {0.02}$. 
%We find that our results are close to the lattice QCD predictions~\cite{Alexandrou:2021mmi}.
Our results are in close agreement with the predictions from lattice QCD~\cite{Alexandrou:2021mmi}.

%==========================================================
\section{Kaon-nucleus induced  Drell-Yan cross section}
%========================================================
The Drell-Yan (DY) process~\cite{Drell:1970wh} has proven to be an effective method for probing the internal structure of hadrons. Over the past few decades, it has helped us explore the structure of not only the proton but also the pion. In the near future, the COMPASS++/AMBER experiment at CERN will measure the DY cross-section of a kaon beam striking a carbon target, which contains equal numbers of protons and neutrons, aiming to explore the structure of the kaon~\cite{Adams:2018pwt}. Here, we predict the cross section for the kaon-carbon induced DY process
with the obtained kaon PDFs supplemented by the PDFs of the target carbon.

%=========================================
\begin{figure}
% 	\begin{center}
			% \includegraphics[width=0.97\linewidth]{kc_msmf_m2.pdf}	\\		\includegraphics[width=0.97\linewidth]{kc_msmf_tau2.pdf}
   \includegraphics[width=0.97\linewidth]{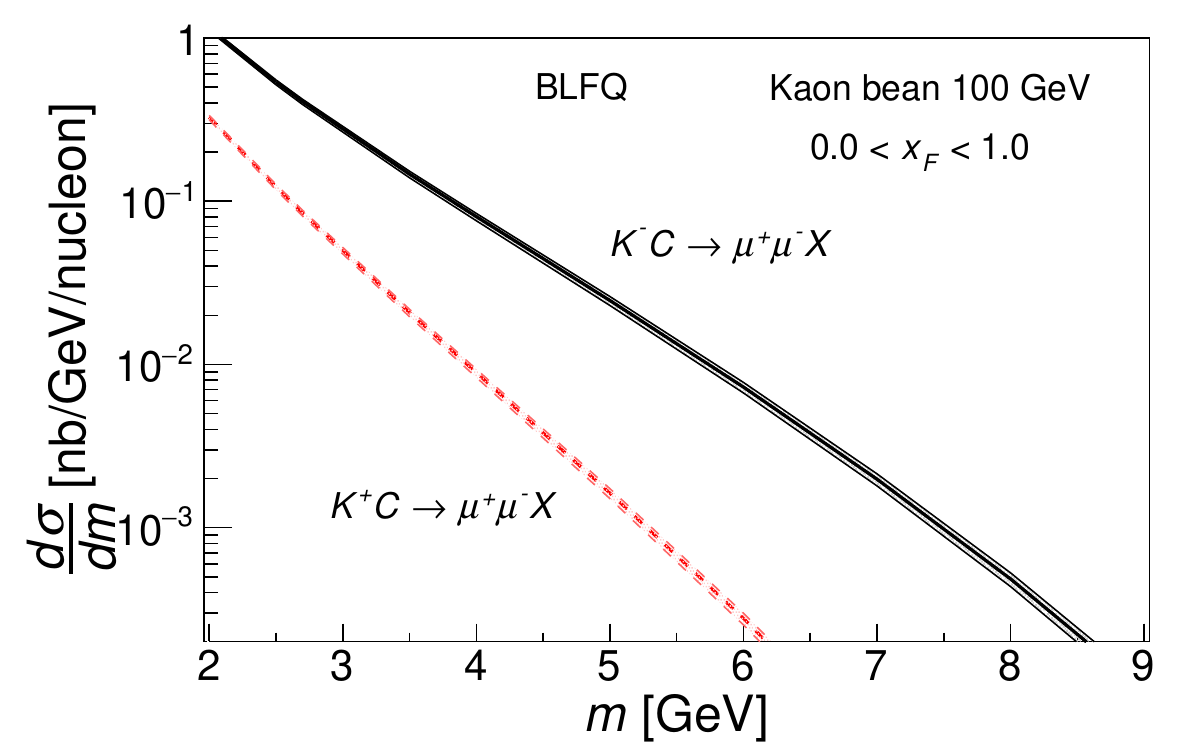}	\\		\includegraphics[width=0.97\linewidth]{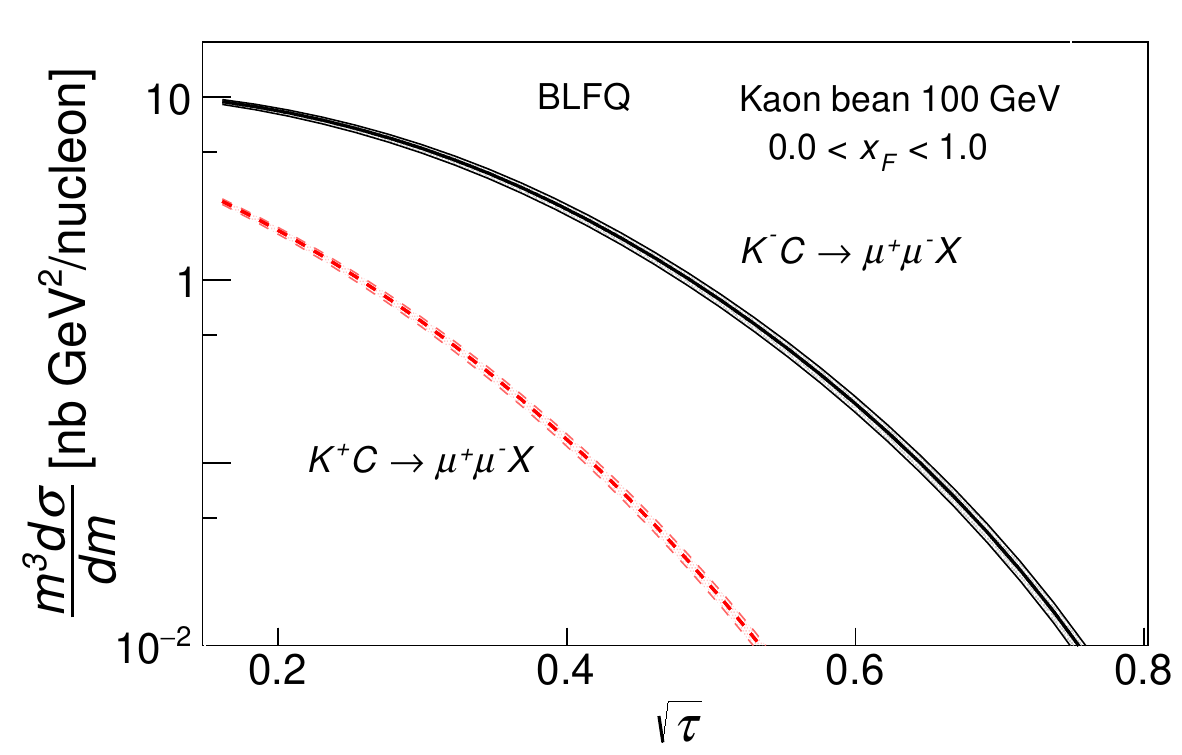}
		\caption{The differential cross section for the kaon-carbon induced DY process. Upper panel: ${\rm d}\sigma /{\rm d}m$ as function of the variable $m$.  Lower panel: $m^3{\rm d}\sigma /{\rm d}m$ as function of the variable $\sqrt{\tau}$. The black and red curves correspond to the results for $K^-$ and $K^+$ beams, respectively. 
        % The error bands reflect a 50\% uncertainty in the normalization factor `$K$'.
        The error bands reflect a $\pm10$\% uncertainty in the initial scale $\mu_0$.
        }
		\label{fig_cross}
% 	\end{center}
\end{figure}
%=========================================

The differential cross section for the kaon-carbon induced DY process, $K^\pm C \to \mu^+\mu^-X$, is given by \cite{Lan:2019rba}
\begin{align}
    \frac{m^3{\rm d}^2\sigma}{{\rm d}m{\rm d}Y}&=\frac{8\pi \alpha_s^2}{9}\frac{m^2}{s}\sum_{ij}\int {\rm d}x_1{\rm d}x_2 \nonumber\\
    & \times \Tilde{C}_{ij}(x_1,x_2,s,m,\mu^2)f^K_{i}(x_1,\mu^2)f^C_{j}(x_2,\mu^2),
    \label{eqn:DYCS_NLO}
\end{align}
where $\widetilde{C}_{ij}$ are the hard-scattering kernels, expandable as a series in the strong coupling $\alpha_s$. At next-to-leading order (NLO), the channels $(ij) = (\bar{q}q), (q\bar{q}), (gq), (qg), (\bar{q}g), (g\bar{q})$ contribute, where partons $i$ and $j$ originate from the kaon and carbon, respectively.
Here, $m$ defines the invariant mass of the lepton pair, $s$ is the center of mass energy square,  $Y$ is the rapidity, $\tau=m^2/s$, and $x_F=x_1-x_2$. To compute the DY cross section in Eq.~(\ref{eqn:DYCS_NLO}), we adopt the carbon PDFs from the nCTEQ 2015~\cite{Kovarik:2015cma} analysis at the experimental scale $\mu^2=16$ GeV$^2$ in conjunction with our kaon PDFs at the same scale.

After integrating out $Y$--dependence of the differential cross section $m^3{\rm d}^2\sigma/{\rm d}m{\rm d}Y$, our predictions are shown in Fig.~\ref{fig_cross} as functions of $m$ and $\sqrt{\tau}$. The black lines represent the results for a carbon target bombarded by a $K^-$ beam with 100 GeV energy~\cite{Adams:2018pwt}, while the red lines correspond to a $K^+$ beam with the same energy. The error bands reflect a $\pm10$\% uncertainty in the initial scale $\mu_0$. We find that the cross-section for $K^-C\to \mu^+\mu^-X$ is significantly larger than that for $K^+C\to \mu^+\mu^-X$. We expect these results to be measured soon by COMPASS++/AMBER.

%==================================
\section{Conclusion and outlook}
%==================================
We have solved the light-front QCD Hamiltonian for light strange mesons by considering them within the constituent quark-antiquark and quark-antiquark-gluon Fock spaces, along with a three-dimensional confinement in the leading Fock sector. The parameters in this Hamiltonian were adopted from the light unflavored mesons~\cite{Lan:2021wok}. The strange quark masses,  $m_s$ and $m_{f_s}$, are the only adjustable parameters, which were fixed to simultaneously reproduce the experimental masses of the $\phi(1020)$ and $f_1(1420)$. The eigenvalues of the Hamiltonian in BLFQ provide a reasonable description of the strangeonia and strange meson mass spectra. The light-front wave functions (LFWFs), obtained as the eigenvectors of this Hamiltonian, were then used to calculate the kaon electromagnetic form factor (EMFF), decay constant, and initial PDFs. We achieved good agreement with experimental data for the kaon EMFF and predicted its decay constant, $f_K^{\rm{BLFQ}}=156.9$ MeV, which closely matches the experimental value of $f_K^{\rm{Exp.}}=155.6\pm 0.4$ MeV~\cite{ParticleDataGroup:2022pth}.

The PDFs at a higher experimental scale were computed using the NNLO DGLAP equations, yielding reasonable agreement with experimental data for the ratio of the up quark distribution in the kaon to that in the pion,  $u_v^K/u_v^{\pi}$ . In this comparison, we used the pion PDF from our previous study, which employed the same BLFQ approach~\cite{Lan:2021wok}. At the experimental scale of $ \mu^2 = 20~{\rm GeV^2} $, we found that the kaon valence up quark PDF falls off at large $x$ as $(1-x)^{2.24}$. At this scale, the valence up quark, valence strange antiquark, sea quarks, and gluons carry approximately 21\%, 26\%, 11\%, and 42\% of the kaon momentum, respectively. Additionally, we have predicted the cross sections for the kaon-carbon-induced Drell-Yan process at NLO, which is expected to be measured soon by COMPASS++/AMBER.

The resulting LFWFs can be utilized to explore various quark and gluon distributions, such as generalized parton distributions, transverse momentum-dependent parton distributions, and double parton correlations in light strange mesons. Additionally, this work can be expanded to higher Fock sectors to include, for instance, sea quark degrees of freedom.

\section*{Acknowledgements}
We thank Jiatong Wu for useful discussions. 
J. Lan is supported by Special Research Assistant Funding Project, Chinese Academy of Sciences, by the National Natural Science Foundation of China under Grant No. 12305095, and the Natural Science Foundation of Gansu Province, China, Grant No. 23JRRA631.
Z. Zhu is supported by the Natural Science Foundation of Gansu Province, China, Grant No. 23JRRA571.
C. Mondal. is supported by new faculty start up funding the Institute of Modern Physics, Chinese Academy of Sciences, Grants No. E129952YR0. 
X. Zhao is supported by new faculty startup funding by the Institute of Modern Physics, Chinese Academy of Sciences, by Key Re- search Program of Frontier Sciences, Chinese Academy of Sciences, Grant No. ZDBS-LY-7020, by the Natural Science Foundation of Gansu Province, China, Grant No. 20JR10RA067, by the Foundation for Key Talents of Gansu Province, by the Central Funds Guiding the Local Science and Technology Development of Gansu Province, Grant No. 22ZY1QA006, by international partnership program of the Chinese Academy of Sciences, Grant No. 016GJHZ2022103FN, by the Strategic Priority Research Program of the Chinese Academy of Sciences, Grant No. XDB34000000, and by the National Natural Science Foundation of China under Grant No.12375143.
J. P. Vary. is supported by the Department of Energy under Grant No. DE-SC0023692. This research is supported by Gansu International Collaboration and Talents Recruitment Base of Particle Physics (2023–2027). A portion of the computational resources were also provided by Taiyuan Advanced Computing Center.

\biboptions{sort&compress}
\bibliographystyle{elsarticle-num}
\bibliography{ref.bib}
\end{document}

%% file: preamble_math_physics.tex
\usepackage{amsmath,amssymb,mathtools} 
\ifpdftex
    \usepackage{bbm}

\else
    \usepackage[bold-style=ISO]{unicode-math}
\fi

\usepackage{siunitx}  
\makeatletter
\newcommand{\vast}[1]{\bBigg@{#1}}
\makeatother

\newcommand{\intd}[3][]{\ifthenelse{\isempty{#3}}{\mathrm{d}^{#1} #2}{\frac{\mathrm{d}^{#1} #2}{#3}}\;}

\usepackage{stackengine}

\newcommand{\cond}{\; | \;}
\newcommand{\setcond}[2]{\ifthenelse{\isempty{#2}}{\{#1\}}{\{#1\cond{}#2\}}}

\usepackage{braket}
\usepackage{cancel}

%% file: MesonPLB.bbl
\begin{thebibliography}{10}
\expandafter\ifx\csname url\endcsname\relax
  \def\url#1{\texttt{#1}}\fi
\expandafter\ifx\csname urlprefix\endcsname\relax\def\urlprefix{URL }\fi
\expandafter\ifx\csname href\endcsname\relax
  \def\href#1#2{#2} \def\path#1{#1}\fi

\bibitem{Rochester:1947mi}
G.~D. Rochester, C.~C. Butler, {Evidence for the Existence of New Unstable
  Elementary Particles}, Nature 160 (1947) 855--857.
\newblock \href {https://doi.org/10.1038/160855a0}
  {\path{doi:10.1038/160855a0}}.

\bibitem{Gell-Mann:1964ewy}
M.~Gell-Mann, {A Schematic Model of Baryons and Mesons}, Phys. Lett. 8 (1964)
  214--215.
\newblock \href {https://doi.org/10.1016/S0031-9163(64)92001-3}
  {\path{doi:10.1016/S0031-9163(64)92001-3}}.

\bibitem{Woods:1988za}
M.~Woods, et~al., {First Result on a New Measurement of $\epsilon^{\prime} /
  \epsilon$ in the Neutral Kaon System}, Phys. Rev. Lett. 60 (1988) 1695.
\newblock \href {https://doi.org/10.1103/PhysRevLett.60.1695}
  {\path{doi:10.1103/PhysRevLett.60.1695}}.

\bibitem{NA31:1993tha}
G.~D. Barr, et~al., {A New measurement of direct CP violation in the neutral
  kaon system}, Phys. Lett. B 317 (1993) 233--242.
\newblock \href {https://doi.org/10.1016/0370-2693(93)91599-I}
  {\path{doi:10.1016/0370-2693(93)91599-I}}.

\bibitem{E731:1993niw}
L.~K. Gibbons, et~al., {Measurement of the CP violation parameter
  Re($\epsilon^{\prime} / \epsilon$)}, Phys. Rev. Lett. 70 (1993) 1203--1206.
\newblock \href {https://doi.org/10.1103/PhysRevLett.70.1203}
  {\path{doi:10.1103/PhysRevLett.70.1203}}.

\bibitem{Amendolia:1986ui}
S.~R. Amendolia, et~al., {A Measurement of the Kaon Charge Radius}, Phys. Lett.
  B 178 (1986) 435--440.
\newblock \href {https://doi.org/10.1016/0370-2693(86)91407-3}
  {\path{doi:10.1016/0370-2693(86)91407-3}}.

\bibitem{Dally:1980dj}
E.~B. Dally, et~al., {DIRECT MEASUREMENT OF THE NEGATIVE KAON FORM-FACTOR},
  Phys. Rev. Lett. 45 (1980) 232--235.
\newblock \href {https://doi.org/10.1103/PhysRevLett.45.232}
  {\path{doi:10.1103/PhysRevLett.45.232}}.

\bibitem{Carmignotto:2018uqj}
M.~Carmignotto, et~al., {Separated Kaon Electroproduction Cross Section and the
  Kaon Form Factor from 6 GeV JLab Data}, Phys. Rev. C 97~(2) (2018) 025204.
\newblock \href {http://arxiv.org/abs/1801.01536} {\path{arXiv:1801.01536}},
  \href {https://doi.org/10.1103/PhysRevC.97.025204}
  {\path{doi:10.1103/PhysRevC.97.025204}}.

\bibitem{Arrington:2021biu}
J.~Arrington, et~al., {Revealing the structure of light pseudoscalar mesons at
  the electron\textendash{}ion collider}, J. Phys. G 48~(7) (2021) 075106.
\newblock \href {http://arxiv.org/abs/2102.11788} {\path{arXiv:2102.11788}},
  \href {https://doi.org/10.1088/1361-6471/abf5c3}
  {\path{doi:10.1088/1361-6471/abf5c3}}.

\bibitem{Chen:2020ijn}
X.~Chen, F.-K. Guo, C.~D. Roberts, R.~Wang, {Selected Science Opportunities for
  the EicC}, Few Body Syst. 61~(4) (2020) 43.
\newblock \href {http://arxiv.org/abs/2008.00102} {\path{arXiv:2008.00102}},
  \href {https://doi.org/10.1007/s00601-020-01574-0}
  {\path{doi:10.1007/s00601-020-01574-0}}.

\bibitem{Proceedings:2020fyd}
S.~J. Brodsky, et~al., {Strong QCD from Hadron Structure Experiments}: {Newport
  News, VA, USA, November 4-8, 2019}, Int. J. Mod. Phys. E 29~(08) (2020)
  2030006.
\newblock \href {http://arxiv.org/abs/2006.06802} {\path{arXiv:2006.06802}},
  \href {https://doi.org/10.1142/S0218301320300064}
  {\path{doi:10.1142/S0218301320300064}}.

\bibitem{Conway:1989fs}
J.~S. Conway, et~al., {Experimental Study of Muon Pairs Produced by 252-GeV
  Pions on Tungsten}, Phys. Rev. D 39 (1989) 92--122.
\newblock \href {https://doi.org/10.1103/PhysRevD.39.92}
  {\path{doi:10.1103/PhysRevD.39.92}}.

\bibitem{NA3:1983ejh}
J.~Badier, et~al., {Experimental Determination of the pi Meson Structure
  Functions by the Drell-Yan Mechanism}, Z. Phys. C 18 (1983) 281.
\newblock \href {https://doi.org/10.1007/BF01573728}
  {\path{doi:10.1007/BF01573728}}.

\bibitem{Adams:2018pwt}
B.~Adams, et~al., {Letter of Intent: A New QCD facility at the M2 beam line of
  the CERN SPS (COMPASS++/AMBER)} (8 2018).
\newblock \href {http://arxiv.org/abs/1808.00848} {\path{arXiv:1808.00848}}.

\bibitem{Accardi:2012qut}
A.~Accardi, et~al., {Electron Ion Collider: The Next QCD Frontier}:
  {Understanding the glue that binds us all}, Eur. Phys. J. A 52~(9) (2016)
  268.
\newblock \href {http://arxiv.org/abs/1212.1701} {\path{arXiv:1212.1701}},
  \href {https://doi.org/10.1140/epja/i2016-16268-9}
  {\path{doi:10.1140/epja/i2016-16268-9}}.

\bibitem{AbdulKhalek:2021gbh}
R.~Abdul~Khalek, et~al., {Science Requirements and Detector Concepts for the
  Electron-Ion Collider}: {EIC Yellow Report}, Nucl. Phys. A 1026 (2022)
  122447.
\newblock \href {http://arxiv.org/abs/2103.05419} {\path{arXiv:2103.05419}},
  \href {https://doi.org/10.1016/j.nuclphysa.2022.122447}
  {\path{doi:10.1016/j.nuclphysa.2022.122447}}.

\bibitem{Anderle:2021wcy}
D.~P. Anderle, et~al., {Electron-ion collider in China}, Front. Phys. (Beijing)
  16~(6) (2021) 64701.
\newblock \href {http://arxiv.org/abs/2102.09222} {\path{arXiv:2102.09222}},
  \href {https://doi.org/10.1007/s11467-021-1062-0}
  {\path{doi:10.1007/s11467-021-1062-0}}.

\bibitem{Londergan:1996vh}
J.~T. Londergan, G.~Q. Liu, A.~W. Thomas, {Kaon - nucleus Drell-Yan processes
  and kaon structure functions}, Phys. Lett. B 380 (1996) 393--398.
\newblock \href {http://arxiv.org/abs/hep-ph/9604448}
  {\path{arXiv:hep-ph/9604448}}, \href
  {https://doi.org/10.1016/0370-2693(96)00500-X}
  {\path{doi:10.1016/0370-2693(96)00500-X}}.

\bibitem{Gluck:1997ww}
M.~Gluck, E.~Reya, M.~Stratmann, {Mesonic parton densities derived from
  constituent quark model constraints}, Eur. Phys. J. C 2 (1998) 159--163.
\newblock \href {http://arxiv.org/abs/hep-ph/9711369}
  {\path{arXiv:hep-ph/9711369}}, \href {https://doi.org/10.1007/s100520050130}
  {\path{doi:10.1007/s100520050130}}.

\bibitem{Watanabe:2018qju}
A.~Watanabe, T.~Sawada, C.~W. Kao, {Meson cloud effects on kaon quark
  distribution functions and the SU(3) flavor symmetry}, Nucl. Part. Phys.
  Proc. 300-302 (2018) 121--125.
\newblock \href {http://arxiv.org/abs/1810.04032} {\path{arXiv:1810.04032}},
  \href {https://doi.org/10.1016/j.nuclphysbps.2018.12.021}
  {\path{doi:10.1016/j.nuclphysbps.2018.12.021}}.

\bibitem{Hutauruk:2023ccw}
P.~T.~P. Hutauruk, S.-i. Nam, {Updated analyses of gluon distribution functions
  for the pion and kaon from the gauge-invariant nonlocal chiral quark model}
  (2 2023).
\newblock \href {http://arxiv.org/abs/2302.05566} {\path{arXiv:2302.05566}}.

\bibitem{Hutauruk:2021kej}
P.~T.~P. Hutauruk, S.-i. Nam, {Gluon and valence quark distributions for the
  pion and kaon in nuclear matter}, Phys. Rev. D 105~(3) (2022) 034021.
\newblock \href {http://arxiv.org/abs/2112.05435} {\path{arXiv:2112.05435}},
  \href {https://doi.org/10.1103/PhysRevD.105.034021}
  {\path{doi:10.1103/PhysRevD.105.034021}}.

\bibitem{Roberts:2021nhw}
C.~D. Roberts, D.~G. Richards, T.~Horn, L.~Chang, {Insights into the emergence
  of mass from studies of pion and kaon structure}, Prog. Part. Nucl. Phys. 120
  (2021) 103883.
\newblock \href {http://arxiv.org/abs/2102.01765} {\path{arXiv:2102.01765}},
  \href {https://doi.org/10.1016/j.ppnp.2021.103883}
  {\path{doi:10.1016/j.ppnp.2021.103883}}.

\bibitem{Xu:2018eii}
S.-S. Xu, L.~Chang, C.~D. Roberts, H.-S. Zong, {Pion and kaon valence-quark
  parton quasidistributions}, Phys. Rev. D 97~(9) (2018) 094014.
\newblock \href {http://arxiv.org/abs/1802.09552} {\path{arXiv:1802.09552}},
  \href {https://doi.org/10.1103/PhysRevD.97.094014}
  {\path{doi:10.1103/PhysRevD.97.094014}}.

\bibitem{Nguyen:2011jy}
T.~Nguyen, A.~Bashir, C.~D. Roberts, P.~C. Tandy, {Pion and kaon valence-quark
  parton distribution functions}, Phys. Rev. C 83 (2011) 062201.
\newblock \href {http://arxiv.org/abs/1102.2448} {\path{arXiv:1102.2448}},
  \href {https://doi.org/10.1103/PhysRevC.83.062201}
  {\path{doi:10.1103/PhysRevC.83.062201}}.

\bibitem{Bednar:2018mtf}
K.~D. Bednar, I.~C. Clo\"et, P.~C. Tandy, {Distinguishing Quarks and Gluons in
  Pion and Kaon Parton Distribution Functions}, Phys. Rev. Lett. 124~(4) (2020)
  042002.
\newblock \href {http://arxiv.org/abs/1811.12310} {\path{arXiv:1811.12310}},
  \href {https://doi.org/10.1103/PhysRevLett.124.042002}
  {\path{doi:10.1103/PhysRevLett.124.042002}}.

\bibitem{Liu:2023yuj}
W.-Y. Liu, E.~Shuryak, I.~Zahed, {Hadronic structure on the light-front VII.
  Pions and kaons and their partonic distributions} (2 2023).
\newblock \href {http://arxiv.org/abs/2302.03759} {\path{arXiv:2302.03759}}.

\bibitem{Kaur:2020vkq}
S.~Kaur, N.~Kumar, J.~Lan, C.~Mondal, H.~Dahiya, {Tomography of light mesons in
  the light-cone quark model}, Phys. Rev. D 102~(1) (2020) 014021.
\newblock \href {http://arxiv.org/abs/2002.01199} {\path{arXiv:2002.01199}},
  \href {https://doi.org/10.1103/PhysRevD.102.014021}
  {\path{doi:10.1103/PhysRevD.102.014021}}.

\bibitem{Lan:2019vui}
J.~Lan, C.~Mondal, S.~Jia, X.~Zhao, J.~P. Vary, {Parton Distribution Functions
  from a Light Front Hamiltonian and QCD Evolution for Light Mesons}, Phys.
  Rev. Lett. 122~(17) (2019) 172001.
\newblock \href {http://arxiv.org/abs/1901.11430} {\path{arXiv:1901.11430}},
  \href {https://doi.org/10.1103/PhysRevLett.122.172001}
  {\path{doi:10.1103/PhysRevLett.122.172001}}.

\bibitem{PhysRevD.103.014516}
H.-W. Lin, J.-W. Chen, Z.~Fan, J.-H. Zhang, R.~Zhang,
  \href{https://link.aps.org/doi/10.1103/PhysRevD.103.014516}{Valence-quark
  distribution of the kaon and pion from lattice qcd}, Phys. Rev. D 103 (2021)
  014516.
\newblock \href {https://doi.org/10.1103/PhysRevD.103.014516}
  {\path{doi:10.1103/PhysRevD.103.014516}}.
\newline\urlprefix\url{https://link.aps.org/doi/10.1103/PhysRevD.103.014516}

\bibitem{Alexandrou:2021mmi}
C.~Alexandrou, S.~Bacchio, I.~Clo\"et, M.~Constantinou, K.~Hadjiyiannakou,
  G.~Koutsou, C.~Lauer, {Pion and kaon
  \ensuremath{\langle}x3\ensuremath{\rangle} from lattice QCD and PDF
  reconstruction from Mellin moments}, Phys. Rev. D 104~(5) (2021) 054504.
\newblock \href {http://arxiv.org/abs/2104.02247} {\path{arXiv:2104.02247}},
  \href {https://doi.org/10.1103/PhysRevD.104.054504}
  {\path{doi:10.1103/PhysRevD.104.054504}}.

\bibitem{Salas-Chavira:2021wui}
A.~Salas-Chavira, Z.~Fan, H.-W. Lin, {First glimpse into the kaon gluon parton
  distribution using lattice QCD}, Phys. Rev. D 106~(9) (2022) 094510.
\newblock \href {http://arxiv.org/abs/2112.03124} {\path{arXiv:2112.03124}},
  \href {https://doi.org/10.1103/PhysRevD.106.094510}
  {\path{doi:10.1103/PhysRevD.106.094510}}.

\bibitem{Vary_2010}
J.~P. Vary, H.~Honkanen, J.~Li, P.~Maris, S.~J. Brodsky, A.~Harindranath, G.~F.
  de~Teramond, P.~Sternberg, E.~G. Ng, C.~Yang,
  \href{http://dx.doi.org/10.1103/PhysRevC.81.035205}{Hamiltonian light-front
  field theory in a basis function approach}, Physical Review C 81~(3) (Mar.
  2010).
\newblock \href {https://doi.org/10.1103/physrevc.81.035205}
  {\path{doi:10.1103/physrevc.81.035205}}.
\newline\urlprefix\url{http://dx.doi.org/10.1103/PhysRevC.81.035205}

\bibitem{Honkanen:2010rc}
H.~Honkanen, P.~Maris, J.~P. Vary, S.~J. Brodsky, {Electron in a transverse
  harmonic cavity}, Phys. Rev. Lett. 106 (2011) 061603.
\newblock \href {http://arxiv.org/abs/1008.0068} {\path{arXiv:1008.0068}},
  \href {https://doi.org/10.1103/PhysRevLett.106.061603}
  {\path{doi:10.1103/PhysRevLett.106.061603}}.

\bibitem{Zhao:2014xaa}
X.~Zhao, H.~Honkanen, P.~Maris, J.~P. Vary, S.~J. Brodsky, {Electron g-2 in
  Light-Front Quantization}, Phys. Lett. B 737 (2014) 65--69.
\newblock \href {http://arxiv.org/abs/1402.4195} {\path{arXiv:1402.4195}},
  \href {https://doi.org/10.1016/j.physletb.2014.08.020}
  {\path{doi:10.1016/j.physletb.2014.08.020}}.

\bibitem{Wiecki:2014ola}
P.~Wiecki, Y.~Li, X.~Zhao, P.~Maris, J.~P. Vary, {Basis Light-Front
  Quantization Approach to Positronium}, Phys. Rev. D 91~(10) (2015) 105009.
\newblock \href {http://arxiv.org/abs/1404.6234} {\path{arXiv:1404.6234}},
  \href {https://doi.org/10.1103/PhysRevD.91.105009}
  {\path{doi:10.1103/PhysRevD.91.105009}}.

\bibitem{Hu:2020arv}
Z.~Hu, S.~Xu, C.~Mondal, X.~Zhao, J.~P. Vary, {Transverse structure of electron
  in momentum space in basis light-front quantization}, Phys. Rev. D 103~(3)
  (2021) 036005.
\newblock \href {http://arxiv.org/abs/2010.12498} {\path{arXiv:2010.12498}},
  \href {https://doi.org/10.1103/PhysRevD.103.036005}
  {\path{doi:10.1103/PhysRevD.103.036005}}.

\bibitem{Nair:2022evk}
S.~Nair, C.~Mondal, X.~Zhao, A.~Mukherjee, J.~P. Vary, {Basis light-front
  quantization approach to photon}, Phys. Lett. B 827 (2022) 137005.
\newblock \href {http://arxiv.org/abs/2201.12770} {\path{arXiv:2201.12770}},
  \href {https://doi.org/10.1016/j.physletb.2022.137005}
  {\path{doi:10.1016/j.physletb.2022.137005}}.

\bibitem{Li:2017mlw}
Y.~Li, P.~Maris, J.~P. Vary, {Quarkonium as a relativistic bound state on the
  light front}, Phys. Rev. D 96 (2017) 016022.
\newblock \href {http://arxiv.org/abs/1704.06968} {\path{arXiv:1704.06968}},
  \href {https://doi.org/10.1103/PhysRevD.96.016022}
  {\path{doi:10.1103/PhysRevD.96.016022}}.

\bibitem{Jia:2018ary}
S.~Jia, J.~P. Vary, {Basis light front quantization for the charged light
  mesons with color singlet Nambu\textendash{}Jona-Lasinio interactions}, Phys.
  Rev. C 99~(3) (2019) 035206.
\newblock \href {http://arxiv.org/abs/1811.08512} {\path{arXiv:1811.08512}},
  \href {https://doi.org/10.1103/PhysRevC.99.035206}
  {\path{doi:10.1103/PhysRevC.99.035206}}.

\bibitem{Lan:2019rba}
J.~Lan, C.~Mondal, S.~Jia, X.~Zhao, J.~P. Vary, {Pion and kaon parton
  distribution functions from basis light front quantization and QCD
  evolution}, Phys. Rev. D 101~(3) (2020) 034024.
\newblock \href {http://arxiv.org/abs/1907.01509} {\path{arXiv:1907.01509}},
  \href {https://doi.org/10.1103/PhysRevD.101.034024}
  {\path{doi:10.1103/PhysRevD.101.034024}}.

\bibitem{Lan:2019img}
J.~Lan, C.~Mondal, M.~Li, Y.~Li, S.~Tang, X.~Zhao, J.~P. Vary, {Parton
  Distribution Functions of Heavy Mesons on the Light Front}, Phys. Rev. D
  102~(1) (2020) 014020.
\newblock \href {http://arxiv.org/abs/1911.11676} {\path{arXiv:1911.11676}},
  \href {https://doi.org/10.1103/PhysRevD.102.014020}
  {\path{doi:10.1103/PhysRevD.102.014020}}.

\bibitem{Lan:2021wok}
J.~Lan, K.~Fu, C.~Mondal, X.~Zhao, j.~P. Vary, {Light mesons with one dynamical
  gluon on the light front}, Phys. Lett. B 825 (2022) 136890.
\newblock \href {http://arxiv.org/abs/2106.04954} {\path{arXiv:2106.04954}},
  \href {https://doi.org/10.1016/j.physletb.2022.136890}
  {\path{doi:10.1016/j.physletb.2022.136890}}.

\bibitem{Zhu:2023lst}
Z.~Zhu, Z.~Hu, J.~Lan, C.~Mondal, X.~Zhao, J.~P. Vary, {Transverse structure of
  the pion beyond leading twist with basis light-front quantization}, Phys.
  Lett. B 839 (2023) 137808.
\newblock \href {http://arxiv.org/abs/2301.12994} {\path{arXiv:2301.12994}},
  \href {https://doi.org/10.1016/j.physletb.2023.137808}
  {\path{doi:10.1016/j.physletb.2023.137808}}.

\bibitem{ParticleDataGroup:2022pth}
R.~L. Workman, et~al., {Review of Particle Physics}, PTEP 2022 (2022) 083C01.
\newblock \href {https://doi.org/10.1093/ptep/ptac097}
  {\path{doi:10.1093/ptep/ptac097}}.

\bibitem{Kogut:1969xa}
J.~B. Kogut, D.~E. Soper, {Quantum Electrodynamics in the Infinite Momentum
  Frame}, Phys. Rev. D 1 (1970) 2901--2913.
\newblock \href {https://doi.org/10.1103/PhysRevD.1.2901}
  {\path{doi:10.1103/PhysRevD.1.2901}}.

\bibitem{Brodsky:1997de}
S.~J. Brodsky, H.-C. Pauli, S.~S. Pinsky, {Quantum chromodynamics and other
  field theories on the light cone}, Phys. Rept. 301 (1998) 299--486.
\newblock \href {http://arxiv.org/abs/hep-ph/9705477}
  {\path{arXiv:hep-ph/9705477}}, \href
  {https://doi.org/10.1016/S0370-1573(97)00089-6}
  {\path{doi:10.1016/S0370-1573(97)00089-6}}.

\bibitem{Karmanov:2008br}
V.~A. Karmanov, J.~F. Mathiot, A.~V. Smirnov, {Systematic renormalization
  scheme in light-front dynamics with Fock space truncation}, Phys. Rev. D 77
  (2008) 085028.
\newblock \href {http://arxiv.org/abs/0801.4507} {\path{arXiv:0801.4507}},
  \href {https://doi.org/10.1103/PhysRevD.77.085028}
  {\path{doi:10.1103/PhysRevD.77.085028}}.

\bibitem{Li:2015iaw}
Y.~Li, V.~A. Karmanov, P.~Maris, J.~P. Vary, {Ab Initio Approach to the
  Non-Perturbative Scalar Yukawa Model}, Phys. Lett. B 748 (2015) 278--283.
\newblock \href {http://arxiv.org/abs/1504.05233} {\path{arXiv:1504.05233}},
  \href {https://doi.org/10.1016/j.physletb.2015.07.014}
  {\path{doi:10.1016/j.physletb.2015.07.014}}.

\bibitem{Glazek:1992aq}
S.~D. Glazek, R.~J. Perry, {Special example of relativistic Hamiltonian field
  theory}, Phys. Rev. D 45 (1992) 3740--3754.
\newblock \href {https://doi.org/10.1103/PhysRevD.45.3740}
  {\path{doi:10.1103/PhysRevD.45.3740}}.

\bibitem{Li:2015zda}
Y.~Li, P.~Maris, X.~Zhao, J.~P. Vary, {Heavy Quarkonium in a Holographic
  Basis}, Phys. Lett. B 758 (2016) 118--124.
\newblock \href {http://arxiv.org/abs/1509.07212} {\path{arXiv:1509.07212}},
  \href {https://doi.org/10.1016/j.physletb.2016.04.065}
  {\path{doi:10.1016/j.physletb.2016.04.065}}.

\bibitem{Brodsky:2014yha}
S.~J. Brodsky, G.~F. de~Teramond, H.~G. Dosch, J.~Erlich, {Light-Front
  Holographic QCD and Emerging Confinement}, Phys. Rept. 584 (2015) 1--105.
\newblock \href {http://arxiv.org/abs/1407.8131} {\path{arXiv:1407.8131}},
  \href {https://doi.org/10.1016/j.physrep.2015.05.001}
  {\path{doi:10.1016/j.physrep.2015.05.001}}.

\bibitem{Vary:2009gt}
J.~P. Vary, H.~Honkanen, J.~Li, P.~Maris, S.~J. Brodsky, A.~Harindranath, G.~F.
  de~Teramond, P.~Sternberg, E.~G. Ng, C.~Yang, {Hamiltonian light-front field
  theory in a basis function approach}, Phys. Rev. C 81 (2010) 035205.
\newblock \href {http://arxiv.org/abs/0905.1411} {\path{arXiv:0905.1411}},
  \href {https://doi.org/10.1103/PhysRevC.81.035205}
  {\path{doi:10.1103/PhysRevC.81.035205}}.

\bibitem{Zhao:2013cma}
X.~Zhao, A.~Ilderton, P.~Maris, J.~P. Vary, {Scattering in Time-Dependent Basis
  Light-Front Quantization}, Phys. Rev. D 88 (2013) 065014.
\newblock \href {http://arxiv.org/abs/1303.3273} {\path{arXiv:1303.3273}},
  \href {https://doi.org/10.1103/PhysRevD.88.065014}
  {\path{doi:10.1103/PhysRevD.88.065014}}.

\bibitem{Xu:2021wwj}
S.~Xu, C.~Mondal, J.~Lan, X.~Zhao, Y.~Li, J.~P. Vary, {Nucleon structure from
  basis light-front quantization}, Phys. Rev. D 104~(9) (2021) 094036.
\newblock \href {http://arxiv.org/abs/2108.03909} {\path{arXiv:2108.03909}},
  \href {https://doi.org/10.1103/PhysRevD.104.094036}
  {\path{doi:10.1103/PhysRevD.104.094036}}.

\bibitem{Kaur:2024iwn}
S.~Kaur, J.~Wu, Z.~Hu, J.~Lan, C.~Mondal, X.~Zhao, J.~P. Vary, {Quark and gluon
  distributions in \ensuremath{\rho}-meson from basis light-front
  quantization}, Phys. Lett. B 851 (2024) 138563.
\newblock \href {http://arxiv.org/abs/2401.03480} {\path{arXiv:2401.03480}},
  \href {https://doi.org/10.1016/j.physletb.2024.138563}
  {\path{doi:10.1016/j.physletb.2024.138563}}.

\bibitem{MartinContreras:2021yfz}
M.~A. Martin~Contreras, E.~Folco~Capossoli, D.~Li, A.~Vega, H.~Boschi-Filho,
  {Pion form factor from an AdS deformed background}, Nucl. Phys. B 977 (2022)
  115726.
\newblock \href {http://arxiv.org/abs/2104.04640} {\path{arXiv:2104.04640}},
  \href {https://doi.org/10.1016/j.nuclphysb.2022.115726}
  {\path{doi:10.1016/j.nuclphysb.2022.115726}}.

\bibitem{Brodsky:2007hb}
S.~J. Brodsky, G.~F. de~Teramond, {Light-Front Dynamics and AdS/QCD
  Correspondence: The Pion Form Factor in the Space- and Time-Like Regions},
  Phys. Rev. D 77 (2008) 056007.
\newblock \href {http://arxiv.org/abs/0707.3859} {\path{arXiv:0707.3859}},
  \href {https://doi.org/10.1103/PhysRevD.77.056007}
  {\path{doi:10.1103/PhysRevD.77.056007}}.

\bibitem{Lepage:1980fj}
G.~P. Lepage, S.~J. Brodsky, {Exclusive Processes in Perturbative Quantum
  Chromodynamics}, Phys. Rev. D 22 (1980) 2157.
\newblock \href {https://doi.org/10.1103/PhysRevD.22.2157}
  {\path{doi:10.1103/PhysRevD.22.2157}}.

\bibitem{Lepage:1979zb}
G.~P. Lepage, S.~J. Brodsky, {Exclusive Processes in Quantum Chromodynamics:
  Evolution Equations for Hadronic Wave Functions and the Form-Factors of
  Mesons}, Phys. Lett. B 87 (1979) 359--365.
\newblock \href {https://doi.org/10.1016/0370-2693(79)90554-9}
  {\path{doi:10.1016/0370-2693(79)90554-9}}.

\bibitem{Efremov:1979qk}
A.~V. Efremov, A.~V. Radyushkin, {Factorization and Asymptotical Behavior of
  Pion Form-Factor in QCD}, Phys. Lett. B 94 (1980) 245--250.
\newblock \href {https://doi.org/10.1016/0370-2693(80)90869-2}
  {\path{doi:10.1016/0370-2693(80)90869-2}}.

\bibitem{Dokshitzer:1977sg}
Y.~L. Dokshitzer, {Calculation of the Structure Functions for Deep Inelastic
  Scattering and e+ e- Annihilation by Perturbation Theory in Quantum
  Chromodynamics.}, Sov. Phys. JETP 46 (1977) 641--653.

\bibitem{Gribov:1972ri}
V.~N. Gribov, L.~N. Lipatov, {Deep inelastic e p scattering in perturbation
  theory}, Sov. J. Nucl. Phys. 15 (1972) 438--450.

\bibitem{Altarelli:1977zs}
G.~Altarelli, G.~Parisi, {Asymptotic Freedom in Parton Language}, Nucl. Phys. B
  126 (1977) 298--318.
\newblock \href {https://doi.org/10.1016/0550-3213(77)90384-4}
  {\path{doi:10.1016/0550-3213(77)90384-4}}.

\bibitem{Salam:2008qg}
G.~P. Salam, J.~Rojo, {A Higher Order Perturbative Parton Evolution Toolkit
  (HOPPET)}, Comput. Phys. Commun. 180 (2009) 120--156.
\newblock \href {http://arxiv.org/abs/0804.3755} {\path{arXiv:0804.3755}},
  \href {https://doi.org/10.1016/j.cpc.2008.08.010}
  {\path{doi:10.1016/j.cpc.2008.08.010}}.

\bibitem{NA3:1980fhh}
J.~Badier, et~al., {Measurement of the $K^- / \pi^-$ Structure Function Ratio
  Using the {Drell-Yan} Process}, Phys. Lett. B 93 (1980) 354--356.
\newblock \href {https://doi.org/10.1016/0370-2693(80)90530-4}
  {\path{doi:10.1016/0370-2693(80)90530-4}}.

\bibitem{Drell:1970wh}
S.~D. Drell, T.-M. Yan, {Massive Lepton Pair Production in Hadron-Hadron
  Collisions at High-Energies}, Phys. Rev. Lett. 25 (1970) 316--320, [Erratum:
  Phys.Rev.Lett. 25, 902 (1970)].
\newblock \href {https://doi.org/10.1103/PhysRevLett.25.316}
  {\path{doi:10.1103/PhysRevLett.25.316}}.

\bibitem{Kovarik:2015cma}
K.~Kovarik, et~al., {nCTEQ15 - Global analysis of nuclear parton distributions
  with uncertainties in the CTEQ framework}, Phys. Rev. D 93~(8) (2016) 085037.
\newblock \href {http://arxiv.org/abs/1509.00792} {\path{arXiv:1509.00792}},
  \href {https://doi.org/10.1103/PhysRevD.93.085037}
  {\path{doi:10.1103/PhysRevD.93.085037}}.

\end{thebibliography}
